\documentstyle[12pt,a4,epsfig,amssymb]{article}

\newcommand{\bea}{\begin{eqnarray}}
\newcommand{\eea}{\end{eqnarray}}
\newcommand{\rp}{\right)}
\newcommand{\lp}{\left(}

\begin{document}

\title{Evidence of discrete scale invariance in DLA
and time-to-failure by canonical averaging}

\author{A. Johansen $^1$ \& D. Sornette $^{2,3}$\\
$^1$ CATS, Niels Bohr Institute, Blegdamsvej 17\\
2100 Copenhagen, Denmark\\
$^2$ Institute of Geophysics and Planetary Physics\\ and
Department of Earth and Space Sciences\\
UCLA, Los Angeles, CA 90095-1567\\
$^3$ Laboratoire de Physique de la Mati\`ere Condens\'ee\\
CNRS UMR6622 and Universit\'e de Nice-Sophia Antipolis,  B.P. 71\\
06108 NICE Cedex 2, France}

\maketitle

\begin{abstract}

{Discrete scale invariance, which corresponds to a partial breaking of the
scaling symmetry, is reflected in the existence of a hierarchy of
characteristic scales $l_0, l_0 \lambda, l_0 \lambda^2,...$ where
$\lambda$ is a preferred scaling ratio and $l_0$ a microscopic cut-off.
Signatures of discrete scale invariance have recently been found in a
variety of systems ranging
from rupture, earthquakes, Laplacian growth phenomena, ``animals'' in
percolation to financial market crashes. We believe it to be a quite general,
albeit subtle phenomenon. Indeed, the practical problem in uncovering an
underlying discrete scale invariance is that standard ensemble averaging
procedures destroy it as if it was pure noise. This is due to the fact, that
while $\lambda$ only depends on the underlying physics, $l_0$ on the contrary
is realisation-dependent.
Here, we adapt and implement a novel so-called ``canonical'' averaging scheme
which re-sets the $l_0$ of different realizations to approximately the same
value. The
method is based on the determination of a realization-dependent effective
critical point obtained from, {\it e.g.}, a maximum susceptibility criterion.
We demonstrate the method on diffusion limited aggregation and a model of
rupture.}

\end{abstract}
\vskip 1cm   in press in Int. J. Mod. Phys. C  (1998)

\pagebreak

\section{Introduction}

With only a finite number $N$ of independent random variables, the first
moment of
the distribution can only be determined with a precision which is of the
order of $1/\sqrt{N}$.
More generally, when introducing interactions between random
variables, small systems will be dominated by fluctuations while it is often
the case that infinite systems have well-defined observables and are so-called
``self-averaging''. As a consequence, phase transitions and critical points
are only well-defined in the infinite size limit. For instance, the critical
value $p_c=0.249$ for simple-cubic bond percolation
is only true for an infinite system \cite{Stauffer}. However, both numerical
simulations and experiments deal with finite realisations or systems. This
means that various quantities of the finite system will obviously be
realization dependent. Thus percolation in finite systems is not simply
characterized by a single number $p_c$ but instead by a distribution of
$p_c$'s belonging to an ensemble of finite realizations generated by the
same statistics \cite{Vanneste}. The situation may be even more delicate in
the presence of some frozen disorder where the infinite size limit may not
be defined uniquely, hence needing a description in terms of a distribution
(``no-self-averaging'') as in spin glasses \cite{Mezard}.

The standard strategy to deal with finite size effects is to average over a
(hopefully large) set of system realizations. In percolation, this means that
for a fixed system size and fixed probability $p$ of bond occupancy, one
generates as many realizations possible and then averages the relevant
quantities over this ensemble of different realizations. However, it has
recently been pointed out that this ``grand-canonical'' \footnote{This term
refers to the fact that the number of bonds is not the same across the
different realizations. This is thus similar to the grand-canonical ensemble
in statistical physics where $p$ in our percolation example plays the role
of an effective chemical potential for the bond occupancy.} ensemble averaging
may hide important physics. This will be the case especially
in disordered systems, due to the introduction of spurious noise of relative
amplitude proportional to the inverse square root of the system volume
\cite{Pazmandi}. The ``grand canonical'' averaging may thus destroy the
information embedded in the fluctuations in finite systems, controlled by
correlation lengths with exponent less than ${2 \over d}$ (${2 \over d}$ is
the minimum value of the correlation length exponent that would not be hidden
by the usual grand canonical averaging \cite{Pazmandi}).

In fact, there is a well-known example which illustrates dramatically how
averaging can destroy the most important physical information
\cite{Souillard} (see
also \cite{Acustica} for a pedagogical introduction).
Consider a wave propagating within a heterogeneous medium. The wave amplitude
obeys the wave hyperbolic equation with quenched random coefficients. The
natural quantity characterizing the propagation is the Green function of this
equation. It turns out that its average over statistically equivalent sample
realizations decays exponentially with the distance from the source with a
characteristic decay length $l_e$ equal to the mean free path. Thus, at
distances from the source larger than $l_e$, the ensemble averaged Green
function is completely smeared out simply because of the destructive
interference between the random phases of different system
realizations, {\it i.e}, the phases are only coherent over a distance
$l_e$ and become incoherent at larger distances. As a consequence,
the ensemble averaging destroys all information on the phases in the
specific realizations and in the same token destroys the information on the
transport
properties at distances larger than $l_e$. Beyond $l_e$, the wave does not
disappear as would be concluded from the examination of the ensemble averaged
Green function, but enters a new diffusive regime characterized by the
{\it non-averaged} Green function, as shown by Anderson \cite{Anderson}.
Realizing this, it is now possible to define a quantity such that its
average describes correctly the diffusive regime beyond $l_e$. This quantity
is the so-called double Green function defined as the product of the advanced
and the retarded Green functions, which is defined in order to ``synchronize''
the configuration dependent phases.

Another example of recent interest is found in the new concept of DNA walks
\cite{Mantegna,DNArneodo}
which shows the relation between the local content of nucleotide pairs
composition along
a DNA molecule. A recent work \cite{Cebrat} demonstrates that the existence
of DNA phases in the
codons, each of which consists of three base-pairs, requires that the
statistical analysis
of DNA walks should be done in the proper DNA phases, respecting positions
in codons.
Otherwise, coding trends of DNA fragments could compensate each other.
This example can be persued using an analogy with language (D. Stauffer,
private communication).
Imagine one wants to understand the use of an alphabet by looking at the
words in
a dictionary. One will not find much relevant information by putting all
the words
together and look for correlations of the letters. However
if ones looks at the words themselves, taking into account the position of
the letters
in each word, then one will find important regularities much more easily.

In order to address a somewhat similar problem in critical phenomena,
Pazmandi {\it et al.} have recently proposed an alternative averaging
procedure to the usual ensemble, or grand-canonical, averaging.
They have coined it ``canonical averaging'' \cite{Pazmandi}, since it
consists of
identifying, for {\em each} system realization, the specific value
of the critical control parameter $p_c^R$. The natural parameter to
study then becomes
\bea
\Delta = \left| p-p_c^R\right|
\eea
instead of $\left| p- \langle p_c \rangle \right|$. $p$ denotes the control
parameter of the system and the averaging is to be over the different
realisations for the same values of $\Delta$. This procedure has been
tested successfully on the Mott-Insulator to Superfluid transition of
interacting bosons with infinite hopping range in a random potential at
zero temperature for which exact renormalization group equations are
available \cite{Pazmandi}. In quenched random Ising and Ashkin-Teller
models, Wiseman and Domany have recently shown in addition that
sample-to-sample fluctuations of the susceptibility maximum measured in
finite systems are smaller by a factor of $70$ than those of the susceptibility
measured in the same finite systems at the special value of the temperature
corresponding to the critical point of an infinite system \cite{Domany}.
This led them to suggest a more efficient simulation method to estimate
exponents from finite size scaling analysis\,: instead of the accepted
procedure of performing simulations for all sizes at one value $p_c$, perform
simulations at various values of $p$ around $p_c$ in order to identify the
sample-dependent maximum of the susceptibility, {\it i.e.}, the
sample-dependent critical value $p_c^R$. They then use the sample-dependent
reduced control $p-p_c^R$ to carry out the averaging.
Similarly, Ballesteros et al.
\cite{Balles} studied the site-diluted Ising model in three dimensions and
 made the important point that only quantities measurable on the same finite
lattice must appear in the finite size scaling Ansatz of observables.

The present paper is concerned with the implementation of these ideas for the
detection of complex exponents and its associated log-periodicity in
critical systems.

The understanding of complex exponents has advanced significantly
\cite{revue,Thesis} in the last few years. Complex exponents reflect a
discrete scale invariance (DSI), {\it i.e.}, the fact that symmetry with
respect to dilation only occurs under magnification under special factors,
which are arbitrary (integer) powers $\lambda^n$ of a preferred scaling ratio
$\lambda$. Complex exponents have been studied in the eighties in relation
to various problems of physics embedded in hierarchical systems.
Only recently, has it been realized that discrete scale invariance and its
associated complex exponents may appear ``spontaneously'' in euclidean
systems and without the need for a pre-existing hierarchy. Systems that
have been found to exhibit self-organized DSI are Laplacian growth models
\cite{PRL,needle}, rupture in heterogeneous systems
\cite{Ball,Anifrani,Sahimi},
animals \cite{SS} (a generalization of percolation), possibly in earthquakes
\cite{SorSam} among other systems \cite{revue}. In addition,
general field theoretical arguments \cite{SS}
indicate that complex exponents are to be expected generically for
out-of-equilibrium and quenched disordered systems.
Log-periodic structures indicate that the system and its underlying physical
mechanisms have characteristic length scale(s). This is extremely
interesting since it may provide important information on the physical
mechanism and hence can be used as a important modelling constraint. Indeed,
simple power law behaviors can apparently be found everywhere, as seen from
the explosion of the concepts of fractals, criticality and
self-organized-criticality. As an example, the power law distribution
of earthquake energies, known as the Gutenberg-Richter law, can be obtained
by many different mechanisms and from a variety of models. Hence, its
usefulness as a modelling constraint is rather doubtful. A power law reflects
the absence of characteristic scales. This is a desirable property in the
quest for universal behaviors but useless for retrieving the underlying
physical
mechanisms and even more so the sample specific signatures useful
for instance for prediction. In contrast, the
presence of log-periodic features may teach us about important physical
structures, which would otherwise be hidden in a description, which is
(falsly) fully scale invariant

A problem however, when identifying log-periodic signatures in some
observable due to an underlying discrete scale invariance, is that the
phase of the
oscillations ``moves'' as the length of the signal over which the averaging
is performed is changed. They thus have an aspect of noise. However, previous
numerical simulations on Laplacian growth models \cite{PRL} and
renormalization group calculations \cite{SS} have taught us that
the presence of real noise modifies the phase in the log-periodic
oscillations in a sample specific way leading to a ``destructive
interference'' between different realisations. This is similar to the
scrambling of the random phases between different system realizations of
a wave propagating in a random system mentioned above. Hence, a first
strategy is to carry out an analysis on each sample realization separately
and {\em then} average over the different realisations, as done in our
previous work on DLA \cite{PRL}. An enticing alternative is to introduce
a new averaging scheme that does not destroy the oscillations, based on the
preceding considerations \cite{Pazmandi}. Our purpose here is thus to adapt
the averaging scheme proposed in \cite{Pazmandi} in terms of a
realization-dependent effective critical point obtained from, {\it e.g.},
a maximum susceptibilty criterion, to the detection of log-periodicity
in Laplacian growth processes and in rupture. The hope is that the use of a
realization-dependent effective critical point should lead
to a ``rephasing'' of the log-oscillations in the averaging process.

We first examine the case of diffusion-limited-aggregation clusters and
the calculation of their complex fractal dimensions. We then turn to
the implementation of the canonical averaging in the time-to-failure
analysis of a dynamical model of rupture.

\section{Laplacian growth models}

Diffusion-limited-aggregation (DLA) is a growth model introduced by Witten
and Sander \cite{Witten} in which particles are introduced one after another
from large distances and from there perform a random walk until they either
reach the perimeter of the growing cluster and stick to it or are removed if
they get too far away from the center. DLA has been much studied as a
paradigm of the spontaneous formation of complex fractal patterns
\cite{Kadanoff}. Here, the fractal (capacity) dimension
is defined by
\bea
M\lp r \rp \propto r^D ,
\eea
where $M\lp r\rp$ is the "mass" of the cluster or, similarly, the number
of particles within a disk of radius $r$ centered on the initial cluster seed.
The numerical value of the fractal dimension of off-lattice
DLA-clusters have been reported to be $D \approx 1.71 \pm 0.02$
\cite{Batchelor,Ossadnik} and $D \approx 1.60 \pm 0.02$ \cite{APRL}
depending on how $D$ is estimated \cite{APRL}. A reason for the
surprisingly large deviation in the dimension reported by different authors
may be the presence of log-periodic oscillations in the local fractal
dimension $D_r$ as a function of $r$, or equivalently $M$, as shown in
\cite{PRL}. Log-periodic oscillations will make estimates of $D$ rather
sensitive to how they are calculated and especially on how the averaging
over the different realisations is performed. A similar phenomenon
\footnote{In ref. \cite{Stausor}, an averaging of the square of the radius
of gyration $r(t)$
 as a function of time $t$ has been performed over many walkers.
In this case, the natural origin of time $t=0$ acts as an effective
``rephasing'' scheme.
This is similar to averaging over DLA cluster, as we perform below, by
counting with
increasing masses and not with increasing radius.} has
been pointed out for the determination of the effective diffusion exponent
versus time of random walks with a fixed bias direction on randomly diluted
cubic lattices above the percolation threshold \cite{Stausor}. The observed
log-periodicity
make it difficult to determine the diffusion exponent for large bias.

Despite their apparent simplicity, a complete analytical theory of Laplacian
growth models is still lacking and most of the understanding
comes from numerical studies. A problem that most theoretical approaches
do not tackle is the existence of a microscopic cut-off, namely the size
of the sticking particles. In other words, has DLA a continuum limit? Our
results showing the presence of log-periodicity suggest that the answer is
negative and that the microscopic scale cascades its way up the hierarchy.
This concept has been explored in simplified Laplacian needle models
\cite{needle},
showing the existence of a cascade of (``ultra-violet'') Mullins-Sekerka
instabilities
that produces an
approximate period-doubling cascade leading to discrete scale invariance
on average \cite{needle}.

We revisit our previous work \cite{PRL} using
a new canonical averaging scheme. 350 large $10^6$ particles off-lattice
DLA clusters were
analysed for log-periodic signatures by constructing the local fractal
dimension as the logarithmic derivative of the mass $M(r)$ of the cluster
{\it versus} the distance from the center $r$,
\bea
D_r \equiv \frac{d\ln M\lp r\rp}{d\ln r}.
\eea
Good and reliable estimates of derivatives of numerical
and experimental data are in general very difficult to obtain \cite{dlanume}
due to ``noise'' {\it etc.} An efficient way of reducing such problems
is to parametrise the data locally by some function and then using
the derivative of {\it that function} as an estimate of the derivative of the
data. Since we are looking for signatures of log-periodic oscillations,
{\it i.e.}, local extrema, we need a function which at least preserves the
second moment of the data. A filter which does this is the Savitzky-Golay
smoothening filter \cite{dlanume} used in \cite{PRL} in order to
obtain an estimate $D_r$ for each cluster.

\begin{figure}
\input{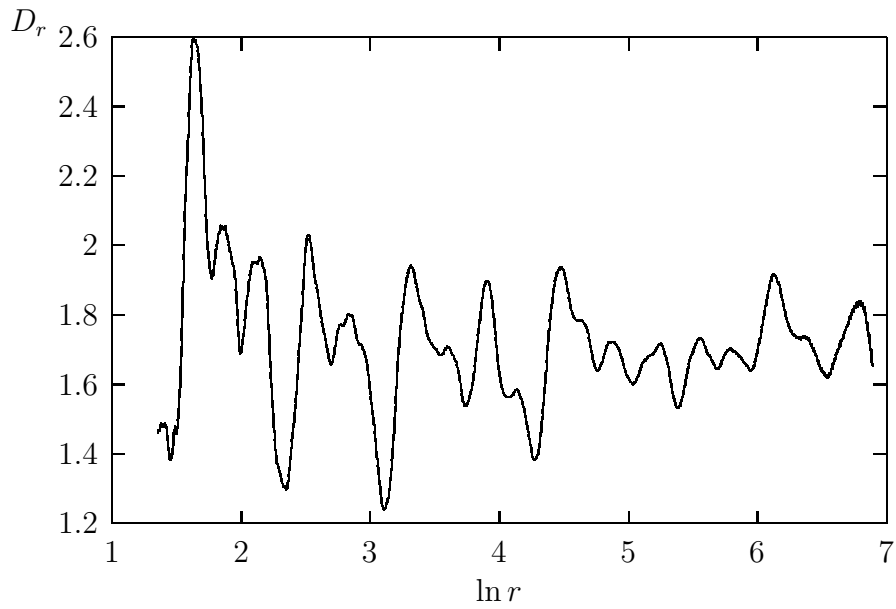}
\caption{\label{sorjoh1} Example of the local dimension $D_r(\log r) =
{{d\log M(r)} \over {d \log r}}$ as a function of $\log r$ for a typical
DLA cluster. The numerical estimate of the  derivative has been obtained
with a Savitsky-Golay smoothing filter.} \end{figure}

A representative example of the local fractal dimension $D_r$ of a DLA cluster
as a function of log-distance to the center of the cluster $\ln r$ is shown in
figure \ref{sorjoh1}. We clearly see that $D_r$  exhibits noisy
quasi-periodic oscillations as
a function of $\ln r$. Furthermore, we see that the
amplitude of the
oscillations decreases as a function of the size of the cluster. In fact,
plots of log-periodic oscillations in the local fractal dimension have
previously been published for both the dielectric breakdown model and the
off-lattice
DLA model as well as on-lattice \cite{Batchelor}, surprisingly enough
without any
comments from the authors. In our previous analysis \cite{PRL}, each of the
350 clusters was analyzed individually in order to estimate the frequencies
of the oscillations. The synthetis of the analysis was to record the
frequency of the two best fits of an oscillating function for each cluster
\cite{PRL}. This is a rather consuming procedure in terms of computing power.
In addition, the significance of the analysis for each cluster is in general
not overwhelming and only the cumulative evidence from the 350 clusters led
to an unambiguous conclusion. Hence, it would be
nice to have an averaging scheme allowing a neater signature.

Upon averaging  an ensemble of DLA clusters with the same radius,
the oscillations disappear (see
for instance figure 3  of Ref.\cite{PRL}). We thus turn to a ``canonical''
averaging scheme.
The ``canonical averaging'' of the 350 clusters has been performed as follows.
First, we need to define a local reduced control parameter that plays the role
of $\Delta = p-p_c^R$ before averaging. In DLA, the critical point
corresponds to the ratio $r/a \to +\infty$ where $r$ is the observation scale
and $a$ is the individual particle size. For a given $r$, different cluster
realizations will have different masses. These mass fluctuations correspond
to the grand-canonical ensemble. We thus propose to carry out the canonical
average by analyzing the local fractal dimension $D$ as a function of the
mass $M$, {\it i.e.} the averaging over the different clusters is to be
performed for the same value of $M$. This transformation is a nonlinear
mapping from $r$ to $M(r)$.

\begin{figure}
\input{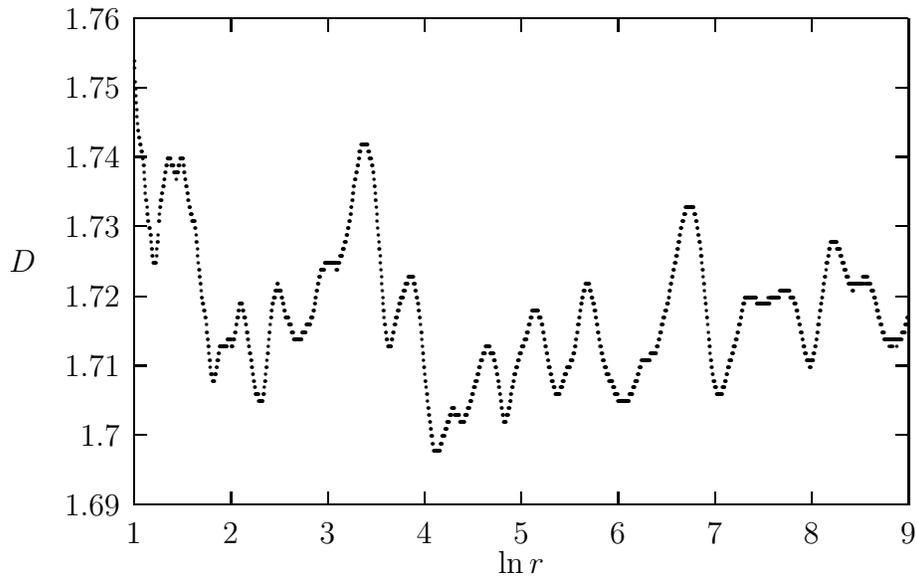}
\caption{\label{sorjoh2} Canonical averaging of the local fractal dimension
of 350 DLA clusters compared to the grand canonical averaging. }
\end{figure}

In fact, this procedure is computationally a very natural choice since $M$ is
the natural counter for the number of particles that have stuck to the cluster.
Roughly speaking, we thus examine $\langle r(M) \rangle$ versus $M$ to find
$1/D(M)$. The local fractal dimension $D$ is thus averaged
over all 350 clusters for the same value of the mass $M\lp r\rp$ by
interpolating between adjacent points in the numerical data. Others have
already
used this averaging procedure (see for instance ref.\cite{Mandel2}). In figure
\ref{sorjoh2}, we see that the amplitude of the log-perodic oscillations has
been somewhat diminished but are still clearly visible.

\begin{figure}
\input{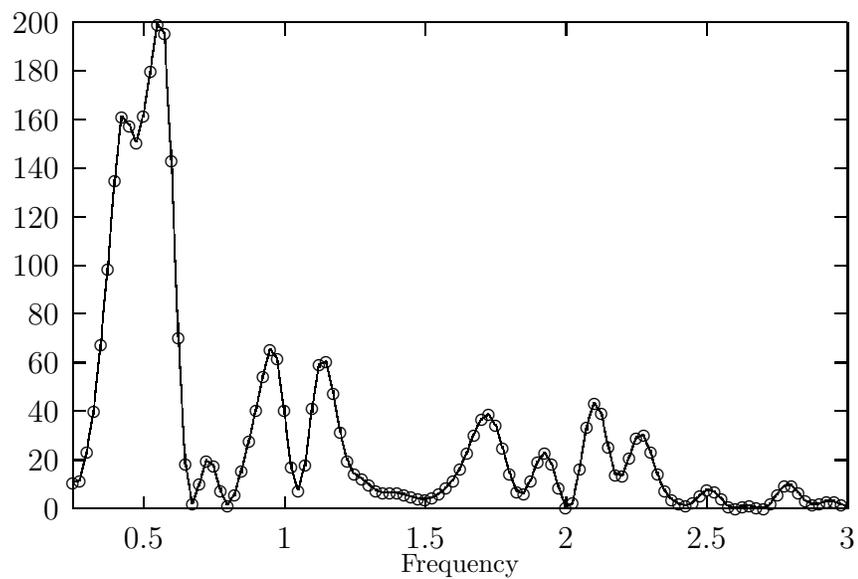}
\caption{\label{sorjoh3} Lomb periodogram analysis of figure
\protect\ref{sorjoh2}.}
\end{figure}

The canonical averaging preserves the oscillations in contrast
to the grand canonical scheme which destroys them completely.
The result from a Lomb periodogram analysis of figure \ref{sorjoh2} is
shown in figure \ref{sorjoh3}. The Lomb periodogram allows us to extract on
small series the main relevant frequencies with a remarkable accuracy (see
\cite{dlanume}). The first peak is very clear and correspond
to the lower frequency ($\approx 0.6$) obtained from the individual
frequency analysis presented in \cite{PRL}. The corresponding preferred scaling
ratio $\lambda$ is approximately $5$. In ref. \cite{PRL}, this peak was
interpreted
as the square $2.3^2$ of the fundamental scaling ratio found around $2.3$.
The second frequency found in
\cite{PRL} ($\approx 1.3$ corresponding to the fundamental scaling
ratio $\lambda =  2.3$) carries some ambiguity since the peak has been split
into two corresponding to a small frequency shift.
Similar results have been
found by averaging over different subsets of the ensemble of 350 clusters.
In particular,
the two main peaks of the Lomb periodogram are robust features.

This novel analysis strengthens the case for log-periodic oscillations in
DLA clusters and more generally in Laplacian growth \cite{needle}. As it
provides a relatively simple procedure to systematically analyze new data,
we hope it will also stimulate other tests.
Note that, in contrast, the available theories (apart from
Ref.\cite{needle}) do not
predict the existence of this discrete scale invariance. The reason is
probably due to not taking into account the microscopic scale cut-off
\cite{Halsey,Pietro}.

The Yale group has presented a series of analysis
\cite{Mandel2,Mandel1,Mandel3}
on fifty one-million particles and twenty ten-million particles clusters, that
stress the importance of non-asymptotic effects  and the existence of a
substantial
amount of sample fluctuations. While it seems delicate, without
a testable theory, to extrapolate
 the observed decrease of the normalized moments and conclude that the data
supports the ``drift'' scenario \cite{Mandel1} of an infinitely continuing
transient,
we note that the logarithmic doubly coordinate figures presented in
\cite{Mandel2}
 exhibit significant systematic {\it oscillations}. In particular,
figure 2 of ref.\cite{Mandel2} of the local dimension as a function of the
logarithm of the particle number, which is similar to our figures 1 and 2,
exhibits very clearly
four log-periodic oscillations with a (modulated) amplitude
significantly larger than the deviation from
zero that the authors are trying to argue. Note that figure 2 of
ref.\cite{Mandel2} is
constructed in the same manner as our figure 2 by averaging clusters with
the same mass.
We concur with ref.\cite{Mandel3} that the mass-lacunarity
effect has to be included, otherwise finite size analysis are bound to be
seriously
distorded. In the light of these and our analysis, it seems more probable
that the most important source of lacunarity is reflected
in a log-periodic behavior, which has a clear physical basis,
relying on a cascade of Mullins-Sekerka period-doubling instabilities
\cite{needle}.

\section{The thermal fuse model} \label{fuse}

Damage is another example of a growth process. Some quasi-static scalar models
of rupture \cite{Niemeyer,Roux} can in fact be put in exact correspondence
with DLA. Here, we do not persue the analysis of the geometry of the
clusters/cracks but turn to the time domain and study the time-dependence of
precursory signals possibly announcing global failure. Time-to-failure
analysis has a long history \cite{Pollock,voigt} and is based on the
detection of an acceleration of some measured signal, for instance acoustic
emissions, on the approach to the global failure. For prediction purposes,
the problem is that an acceleration for instance modelled by a power law
in general provides a poor constraint on the actual time of global rupture.
Log-periodic structures, however, have the potential to improve
significantly the
prediction ability since the parameters of the fitting function can
``lock-in'' on the oscillations to get a more constrained
determination of the global rupture time. This was first tested
on engineering fiber composite structures \cite{Anifrani} for which
log-periodic signatures on the acoustic energy radiated as a
function of time-to-failure could be identified.
Since then, this has been tested extensively on several tens of
pressure tanks made of kevlar-matrix
and carbon-matrix composites constructed by A\'erospatiale Inc., France.
The results from these tests indicate that a precision of
a few percent in the determination of the stress at rupture
is obtained using acoustic emission recorded $20~\%$ below
the stress at rupture. These results have warranted the selection of this
non-destructive
evaluation technique as the routine qualifying procedure in the industrial
fabrication
process.

Here, we would like to present a cleaner analysis that demonstrates the
existence of log-periodicity in the time-to-failure signals. In this goal, we
revisit the thermal fuse model \cite{tfm} (initially introduced in
an electric framework, this model is straightforwardly translated into
its mechanical analog). This is a genuine dynamical model, with
mode III antiplane elasticity,
here put on a square lattice where the bonds are elastic elements with
elastic compliances
$g_i$ uniformly
distributed in the interval $\left[g - \Delta : g + \Delta \right]$. A
force $F$
is applied  across two parallel boundaries of the lattice and
periodic boundary conditions are assumed in the perpendicular direction.
Due to the force, each bond is damaged progressively according a standard
power law
dependence. In addition, a degree of healing is permitted.
The damage variable $D_n$ of bond $n$ varies according to
\bea \label{curtemp}
\frac{d D_n}{d t} = g [f_n \lp t \rp]^b  - a D_n\lp t\rp ,
\label{rtff}
\eea
where the first term of the r.h.s. represent the damage rate due to the force
$f_n$ exerted on this bond and
the second term is the healing term. The limit $b \to \infty$ recovers the
quasi-static model of rupture where the most stressed bond ruptures
\cite{Arcangelis}.
In the simulations used here,
$b=2$, $a=1$ and $g \in \left[ 1\pm 0.9\right]$.
At time $0$, the force $F$ is applied on the system. As a consequence, all
bonds
start to get progressively damaged. The bond whose damage variable reaches
first $1$
breaks down. The force that it supported is then immediately redistributed
according to the
law of (equilibrium) elasticity (long-range Green function), corresponding to
a novel mechanical equilibrium (we thus deal with a so-called quasi-static
model of rupture; the dynamics stems from the damage behavior.). As
a consequence, all the forces in the remaining bonds are modified and their
new values
enter the damage rate equation (\ref{rtff}). A second bond will break and
so on.

\begin{figure}
\input{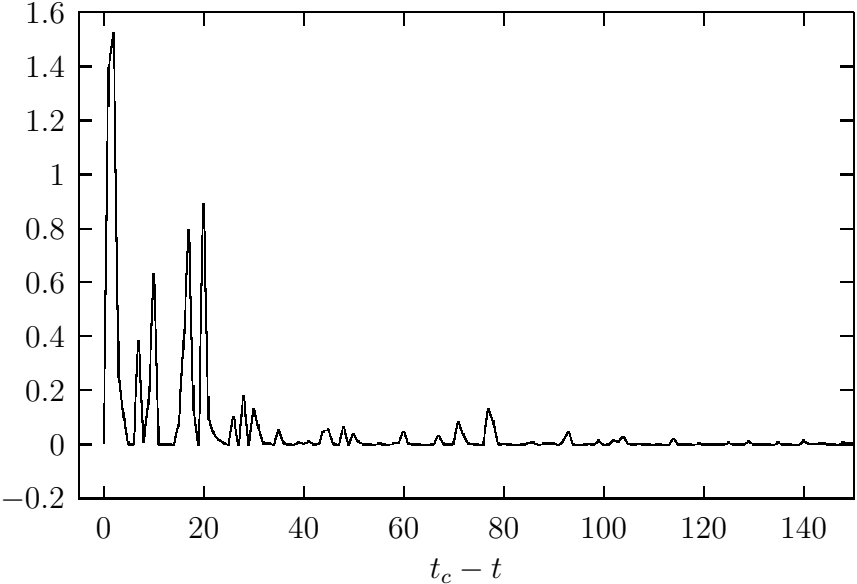}
\caption{\label{sorjoh4} A single simulation of the thermal fuse model.}
\end{figure}

\begin{figure}
\input{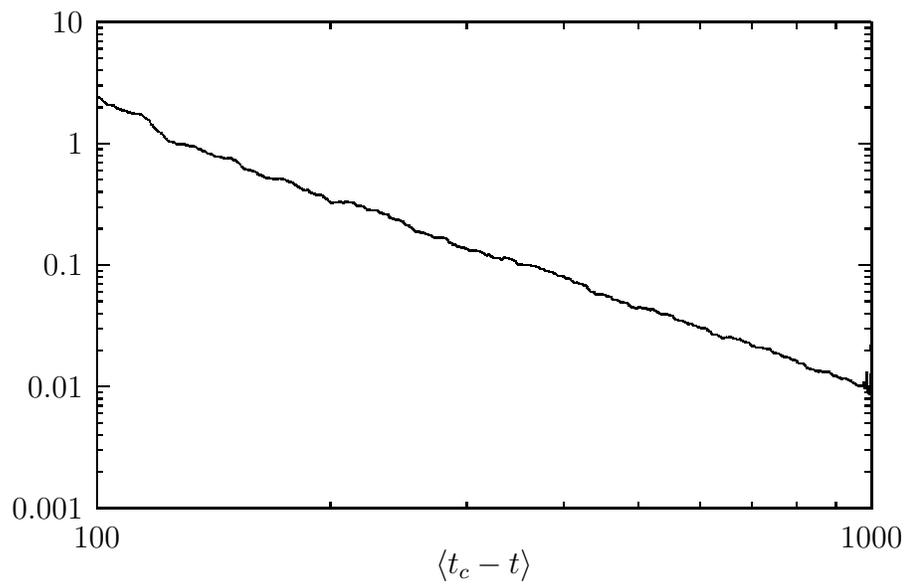}
\caption{\label{sorjoh5} An average of the elastic energy release rate or
rate of broken bonds as a function of time. The average is over 19
simulations of the thermal fuse model,
using the standard (``grand canonical'') averaging scheme, in terms of the time
$t$ from the beginning of the rupture process.}
\end{figure}

An example of a simulation is shown in figure \ref{sorjoh4}, where the last
point corresponds to a complete disconnection or ``total
rupture''. It has previously been established that the ensemble average
of the total dissipated power,
or equivalently of the released elastic energy $E$, behaves as a power law
\bea \label{fusepow}
E \sim \lp t_c - t \rp ^{-\alpha} ,
\eea
as a function of time to rupture $t_c - t$ \cite{tfm}. This result was
obtained through a (grand canonical)
ensemble averaging over different realisations. It has
also been confirmed experimentally on an analogous dynamical electric breakdown
experimental composite system \cite{Lamai}. See figure \ref{sorjoh5} which shows
the average over 19
simulations of the thermal fuse model of the elastic energy release rate or
rate of broken bonds as a function of time. The average is here performed
using the standard (``grand canonical'') averaging scheme, in terms of the time
$t$ from the beginning of the rupture process.

\begin{figure}
\input{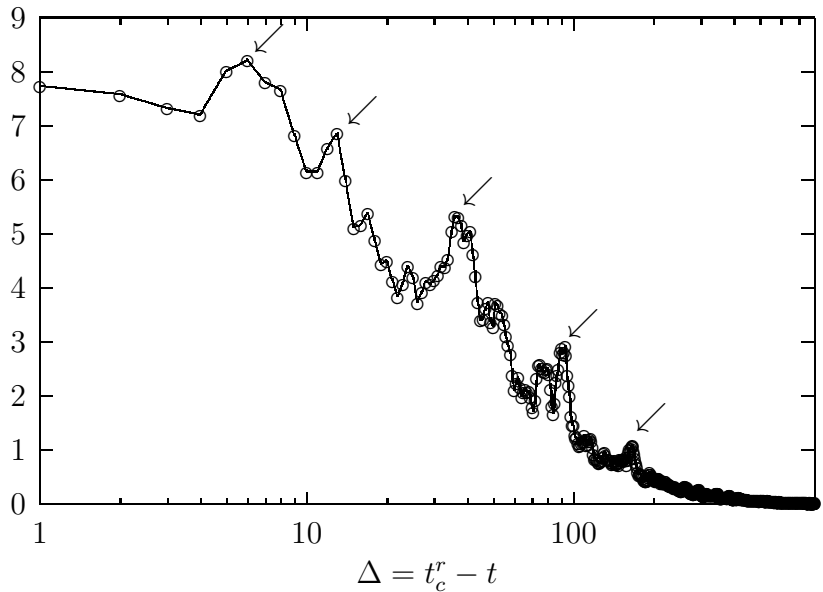}
\caption{\label{sorjoh6} An average of the elastic energy release rate or
rate of broken bonds as a function of time. The average is over the same 19
simulations of the thermal fuse model used in figure \ref{sorjoh5}, but using
the canonical averaging scheme
described in the text, in terms of the effective time-to-failure measured by
the sample specific susceptibility maximum.}
\end{figure}

However, when looking at individual data, we see fluctuations that suggest,
similarly to the experimental pressure tank case, the existence of systematic
log-periodic structures. In order to unravel them, we have re-analysed
the data presented in \cite{tfm} performing a canonical averaging procedure.
Specifically, we identify the time $t_c^r$ for {\it each} disorder
realization, where the second derivative of $E$ (representing a kind of
susceptibility) had its maximum value.  Then, the energy release rate is
averaged over the $19$ available realisations for the same value of $\Delta =
t^r_c - t$. Due to the noisy nature of the data, see figure \ref{sorjoh4},
the derivatives was calculated in the following fashion. First, the cumulative
distribution of broken bonds or released energy as a function of time was
calculated. Then the distribution was lightly smoothed with a moving average
filter with a window of $11$ points  and the second derivative was calculated
using the smoothen data. Before showing the result of the canonically averaged
data, examine the ensemble average of the smoothened data in figure
\ref{sorjoh5}. As previously established in \cite{tfm}, the average energy
release follows a power law $E \sim \lp t_c - t \rp ^{-\alpha}$ as the time
of rupture $T_c$ is approached. Hence, the smoothening of the data has not
in any way added any new and artificial features. For comparison, we see in
figure \ref{sorjoh6} the canonically averaged energy release rate as a
function of $t^r_c - t$. Five approximately equidistant (in
$\ln \lp t^r_c - t\rp$) peaks are visible. This means that the previous
equation
(\ref{fusepow}) should be replaced by
\bea \label{fuselogpow}
E \sim \lp t_c - t \rp ^{-\alpha} P\lp\frac{\ln\lp t_c-t\rp}{\ln\lambda}\rp ,
\eea
where $P$ is a periodic function of period $1$. If we estimate the
corresponding value of
$\lambda$ from the peaks shown in figure \ref{sorjoh6} we get
$\lambda \approx 2.4 \pm 0.4$, {\it i.e.}, again close to 2 as in the case
of DLA. That the DLA model
and the thermal fuse model have approximately the same preferred scaling
ratio is quite reassuring considering the close relationship between the
two models.

\subsection{Conclusion}

We have studied two numerical models, the DLA model and the
thermal fuse model. Instead of the usual ensemble averaging over different
disorder realisations, a novel averaging scheme adapted
to preserve subtle correlated fluctuations in the data have been used with
remarkably success, particularly in the case of the rupture model.

The analysis presented here provides yet another example of spontaneous
generation of discrete scale invariance in a growth model. This
is quite remarkable considering that no pre-existing hierarchy
 exists in the formulation of the model. The discrete scale
invariance embodied in (\ref{fuselogpow}) is created {\em dynamically},
{\it i.e.}
self-organized through the interactions between the cracks.

As to the underlying physical origin of the observed log-periodic
oscillations in DLA and the thermal fuse model, it is probably a cascade
of crack tip instabilities as shown numerically in \cite{needle} with respect
to simplified Laplacian growth models. We find it rather remarkable that
log-periodicity is present in the random rupture model in particular
in view of the fact that cracks nucleate all over the two-dimensional
system in a
disordered manner. However, the long-range nature of elastic interactions
apparently provides a sufficiently strong ordering force.

Finally, our essential message in the light of the recent discussions
\cite{Pazmandi,Domany} and the results presented here is that it would be
fruitful to revisit heterogeneous systems with special attention to the
possible degradation of information induced by standard averaging procedures.
It is our hope that this might stimulate the invention of novel
approaches to the physics of random systems than that offered by the
traditional tool of the field.

Acknowledgement\,:
The authors wish to thank Kristian Schaadt for help with the IDL interpolating
program used in the numerical analysis of the DLA data and Christian
Vanneste for help
in the analysis of the thermal fuse model. We are grateful to D. Stauffer for
very interesting comments on the manuscript that helped clarify and improve
its content.

\newpage
\clearpage

\clearpage

\clearpage

\clearpage

\clearpage

\clearpage

\end{document}